\documentclass[twocolumn,aps,prl,floats,longbibliography]{revtex4-1}
\usepackage{epsfig}
\usepackage{graphicx}
\usepackage[figuresright]{rotating}
\usepackage{relsize}%

\usepackage{amsfonts}
\usepackage{amsmath}
\usepackage{amssymb}
\usepackage{wasysym}
\PassOptionsToPackage{normalem}{ulem}
\usepackage{ulem}
\usepackage[unicode=true,pdfusetitle,
 bookmarks=true,bookmarksnumbered=false,bookmarksopen=false,
 breaklinks=false,pdfborder={0 0 1},backref=false,colorlinks=false]
 {hyperref}
\hypersetup{colorlinks,linkcolor=blue,citecolor=blue,urlcolor=blue}

\makeatletter

\usepackage{xcolor}
\usepackage{pdfcolmk}
\usepackage{wasysym}

\usepackage{txfonts}
 
\makeatother

\begin{document}
\title{Dilemma in strongly correlated materials: \\
Hund's metal vs relativistic Mott insulator}
\author{Gang Chen}
\email{gangchen@hku.hk}
\affiliation{Department of Physics and HKU-UCAS Joint Institute for Theoretical and 
Computational Physics at Hong Kong, The University of Hong Kong, Hong Kong, China}

\begin{abstract} 
We point out the generic competition between the Hund's coupling and the
spin-orbit coupling in correlated materials, and this competition leads to an 
electronic dilemma between the Hund's metal and the relativistic insulators. 
Hund's metals refer to the fate of the would-be insulators where the Hund's 
coupling suppresses the correlation and drives the systems into correlated 
metals. Relativistic Mott insulators refer to the fate of the would-be metals where 
the relativistic spin-orbit coupling enhances the correlation and drives the 
systems into Mott insulators. These contradictory trends are naturally present 
in many correlated materials. We study the competition between Hund's coupling
and spin-orbit coupling in correlated materials and explore the interplay and the 
balance from these two contradictory trends. The system can become 
a {\sl spin-orbit-coupled Hund's metal} or a {\sl Hund's assisted relativistic Mott insulator}. 
Our observation could find a broad application and relevance to many correlated
materials with multiple orbitals. 
\end{abstract}
 
\maketitle

Correlated quantum materials provide a rich platform to explore different 
competing interactions. The simplest one would be the competition between 
the electron kinetic energy and the Coulomb interactions between the electrons. 
This is captured by the well-known Hubbard model~\cite{Hubbard1963}. For the integer electron 
filling per site, a strong on-site electron interaction would directly convert the   
system from a metal into a Mott insulator with the formation of local 
moments~\cite{PhysRev.79.350,PhysRev.100.564,KANAMORI195987}. 
This ``big'' parent picture is decorated in many different ways when extra interactions 
are included or emerge as the subleading effects. This includes, for example, 
the residual interactions and the magnetic ground states of the Mott insulators,
the nature of the metallic states, the band structure topology~\cite{RevModPhys.82.3045,RevModPhys.88.021004}, 
the nature of Mott transition~\cite{PhysRevB.40.546,PhysRevB.78.045109,PhysRev.137.A1726}, 
the orbital selectivity of Mott transition~\cite{PhysRevLett.92.216402,PhysRevLett.110.146402,PhysRevLett.102.126401}, etc. 
Two interesting decorations, Hund's metal~\cite{Georges_2013,medici2017hunds} 
and relativistic Mott insulator~\cite{Witczak_Krempa_2014,2019NatRP...1..264T} 
that are discussed in this work, are from the Hund's coupling and from the spin-orbit coupling, 
respectively. They are two interesting ideas that emerge in the theory of correlated electron 
materials over the past decade.  

The Hund's metal is concerned with how the correlated metallic regime is enhanced
by the presence of the Hund's coupling on multiply degenerate $d$-orbitals
for many transition metal compounds~\cite{Georges_2013,medici2017hunds}. 
Most often, it was argued that, the Hund's coupling effectively reduces the electron correlation
by harnessing the interaction energy gain in the spin sector. Thus, the metallic regime
is significantly expanded compared to the case without the Hund's coupling. 
The relativistic Mott insulator is a concept about the role of strong spin-orbit
coupling in correlated materials with multiple $4d$/$5d$ orbitals or bands~\cite{Witczak_Krempa_2014}. 
It can sometimes be relevant for systems with $3d$ electrons when the 
spin-orbit coupling becomes active. The strong spin-orbit coupling
 twists the 
electron motions as it hops on the lattice and reduces the bandwidth
of the electrons. Another description of this effect is that, the spin-orbit coupling
breaks the whole bands into multiple spin-orbital-entangled subbands with much
narrower bandwidths. As a result, the electron correlation is enhanced. 
The system becomes a spin-orbit-coupled Mott insulator. If there is no spin-orbit coupling,
the system would be a correlated metal. It is the spin-orbit coupling that assists the 
electron correlation and drives the Mott insulating behaviors.
As the spin-orbit coupling is a relativistic effect, the spin-orbit-coupled Mott insulator 
is often referred as the relativistic Mott insulator. 
Having explained the underlying ideas of Hund's metal and relativistic Mott insulator, 
one would immediately realize that, both of these two concepts are dealing with 
the correlated materials with multiple orbitals,
but they have rather opposite tendencies. Thus, the correlated materials would 
fall into a dilemma between Hund's metal and relativistic Mott insulator. 
The purpose of this work is to point out this dilemma through a specific example. 
The specific example in this work is simply adopted to explain the universal physics 
behind, and should not be interpreted as a localized specifics without any generalization. 
One should really extract the general message delivered by our specific example. 
We further design a calculation formalism to study the spin-orbit coupling, 
the Hund's coupling and Mott physics in correlated materials, and hope to
capture these competing effects at least in a crude manner.

\begin{figure}[b]
\includegraphics[width=0.48\textwidth]{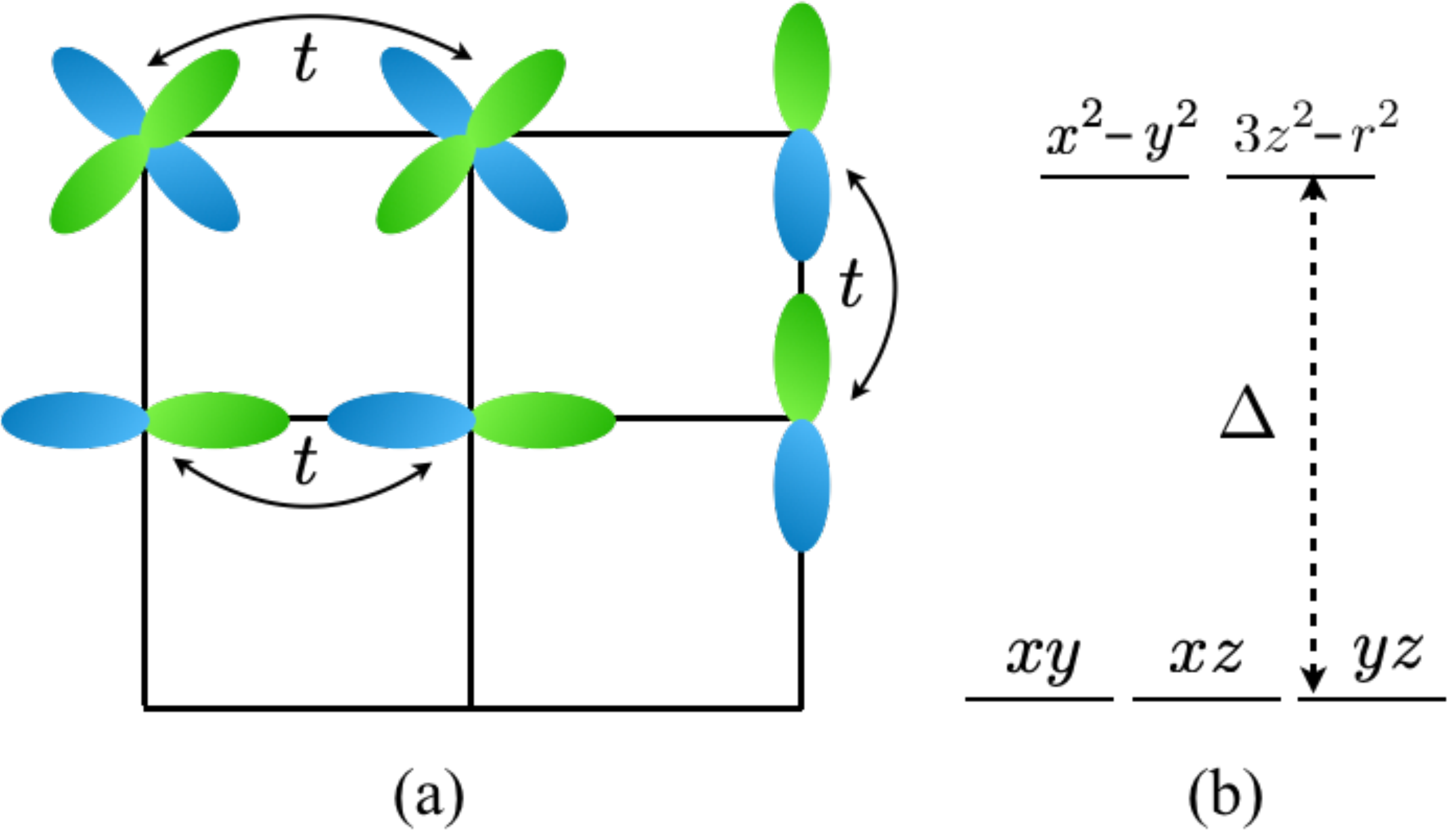}
\caption{(a) The hopping between the $t_{2g}$ orbitals from neighboring sites 
on the square lattice. (b) The energy splitting of three-fold degenerate $t_{2g}$ orbitals 
and the two-fold degenerate $e_g$ orbitals.}
\label{fig1}
\end{figure}

We start with an extended Hubbard model with multiple orbitals on a square lattice (see Fig.~\ref{fig1}). 
We consider the octahedral crystal field environment for the transition metal ions.
The two-fold degenerate $e_g$ orbitals are higher in the energy than the three-fold
degenerate $t_{2g}$ orbitals. We assume that, the crystal field separation $\Delta$ between
the $e_g$ and the $t_{2g}$ orbitals is very high so that we can safely neglect the 
presence of the upper $e_g$ orbitals. We restrict ourselves to the $t_{2g}$ orbitals.
This is sufficient for revealing the physics that was previously advocated. The 
extended Hubbard model on the $t_{2g}$ manifold is written as
\begin{widetext}
\begin{eqnarray}
{\mathcal H} &=& \sum_{\langle ij \rangle} \sum_{m,n}  t_{ij}^{mn} c^\dagger_{im\alpha} c^{}_{jn\alpha}   
+ \sum_i \sum_{m,n} \sum_{\mu}  \frac{\lambda}{2}  {L}^{\mu}_{mn}  {\sigma}^{\mu}_{\alpha\beta} 
c^\dagger_{im\alpha} c^{}_{in\beta}   
+  \sum_i   U  \sum_{m} c^\dagger_{i m\alpha}  c^\dagger_{im\beta} c^{}_{im \beta} c^{}_{im\alpha} 
\nonumber \\
&& +
 \sum_{i} \Big[
 \frac{U'}{2} \sum_{m \neq n}  c^\dagger_{i m\alpha}  c^\dagger_{in\beta} c^{}_{in \beta} c^{}_{im\alpha} 
 + \frac{J}{2}  \sum_{m \neq n} c^\dagger_{i m\alpha}  c^\dagger_{in\beta} c^{}_{im \beta} c^{}_{in\alpha}  
 + \frac{J'}{2}  \sum_{m \neq n} c^\dagger_{i m\alpha}  c^\dagger_{im\beta} c^{}_{in \beta} c^{}_{in\alpha}  
 \Big]  
 + \cdots,
 \label{eq1}
\end{eqnarray}
\end{widetext}
where ${m, n= 1,2,3}$ are the orbital indices corresponding to the $yz, xz, xy$ 
orbitals for the $t_{2g}$ orbitals,  
${\alpha,\beta= \uparrow,\downarrow}$ are the spin indices, ${\mu=x,y,z}$ 
refers to the component for the spin and orbital angular momenta,
and the spin indices are automatically summed. 
The operator $c^\dagger_{im\alpha}$ ($c^{}_{im\alpha}$) creates   
(annihilates) an electron on the $m$ orbital with the spin quantum $\alpha$. 
On the first line of Eq.~\eqref{eq1}, the first term describes the electron 
hopping between different orbitals and the neighboring sites on the lattice,    
the second term describes the atomic spin-orbit coupling, 
and the third term is 
the intra-orbital Coulomb interaction. On the second line of Eq.~\eqref{eq1}, 
the first term is the inter-orbital interaction, the second term is the Hund's coupling, 
the third term describes the electron pair hopping, and ``$\cdots$'' 
refers to the extra interactions and effects that are not considered here. 
Owing to the lattice symmetries and the orbital orientations, 
only the nearest-neighbor intra-orbital hopping is non-vanishing and is set to $t$ 
(see Fig.~\ref{fig1}). 
The interactions in Eq.~\eqref{eq1} are the standard Kanamori interactions. 
We will take the atomic limit with ${J=J'}$ and ${U'=U-2J}$ in our calculation, 
and the interactions can then be described as $U$-interaction and $J$-interactions. 
This Hamiltonian is particularly relevant for e.g.
V$^{4+}$ ions with $3d^1$ electron configurations in Sr$_2$VO$_4$~\cite{PhysRevLett.103.067205,PhysRevLett.95.176405,PhysRevLett.95.176404,PhysRevLett.99.136403}, 
Mo$^{4+}$ ions with $4d^2$ electron configurations in Sr$_2$MoO$_4$~\cite{Ikeda_2000}, 
Re$^{4+}$ ions with $5d^3$ electron configurations in Sr$_2$ReO$_4$, 
Ru$^{4+}$ ions with $4d^4$ electron configurations in Sr$_2$RuO$_4$~\cite{RevModPhys.75.657} and 
Ca$_2$RuO$_4$~\cite{PhysRevB.58.847,PhysRevB.61.R5053,PhysRevB.56.R2916},
even Ir$^{4+}$ ion with $5d^5$ electron configurations in Sr$_2$IrO$_4$ 
after twisting the hoppings~\cite{PhysRevLett.102.017205}, and other transition metal compounds.  
In this work, we analyze the $d^2$ and $d^3$ configurations for the illustration.

Both the Hund's metal and the relativistic Mott insulator are more or less concerned 
with the Mott transition. To tackle with the Mott transition of Eq.~\eqref{eq1}, we implement 
a slave-rotor formulation of the electron operator~\cite{PhysRevB.70.035114}, 
express the electron operator as ${C_{im\alpha} = e^{-i \theta_i} f_{im\alpha}}$, then 
decouple the extended Hubbard model into the spin sector and the charge sector. 
This approximation automatically assumes the spin-charge separation by 
placing the electron charge on the bosonic rotor and placing the spin and
the orbital on the fermionic spinon. Often, this approach was used to 
describe the quantum spin liquid that is proximate to Mott transition~\cite{PhysRevLett.95.036403}. 
Magnetic orders can be incorporated into this formalism by including the 
$J$-interactions onto the spinon sector and properly decoupling the interaction
according to the orderings~\cite{PhysRevB.83.134515}. 
The latter approach attributes the Mott localization
to the $U$-interaction, and the magnetic orders to the $J$-interactions, even 
though both interactions together give rise to the magnetism in many cases.
Here we are not interested in addressing the nature of the ground state for the 
Mott regime, but to understand the variation of the Mott transition.  Thus, we 
simply assume the Mott side is a spin liquid and explore the fate of Mott transition
in the presence of extra couplings.

\begin{figure*}[t]
\includegraphics[width=0.238\textwidth]{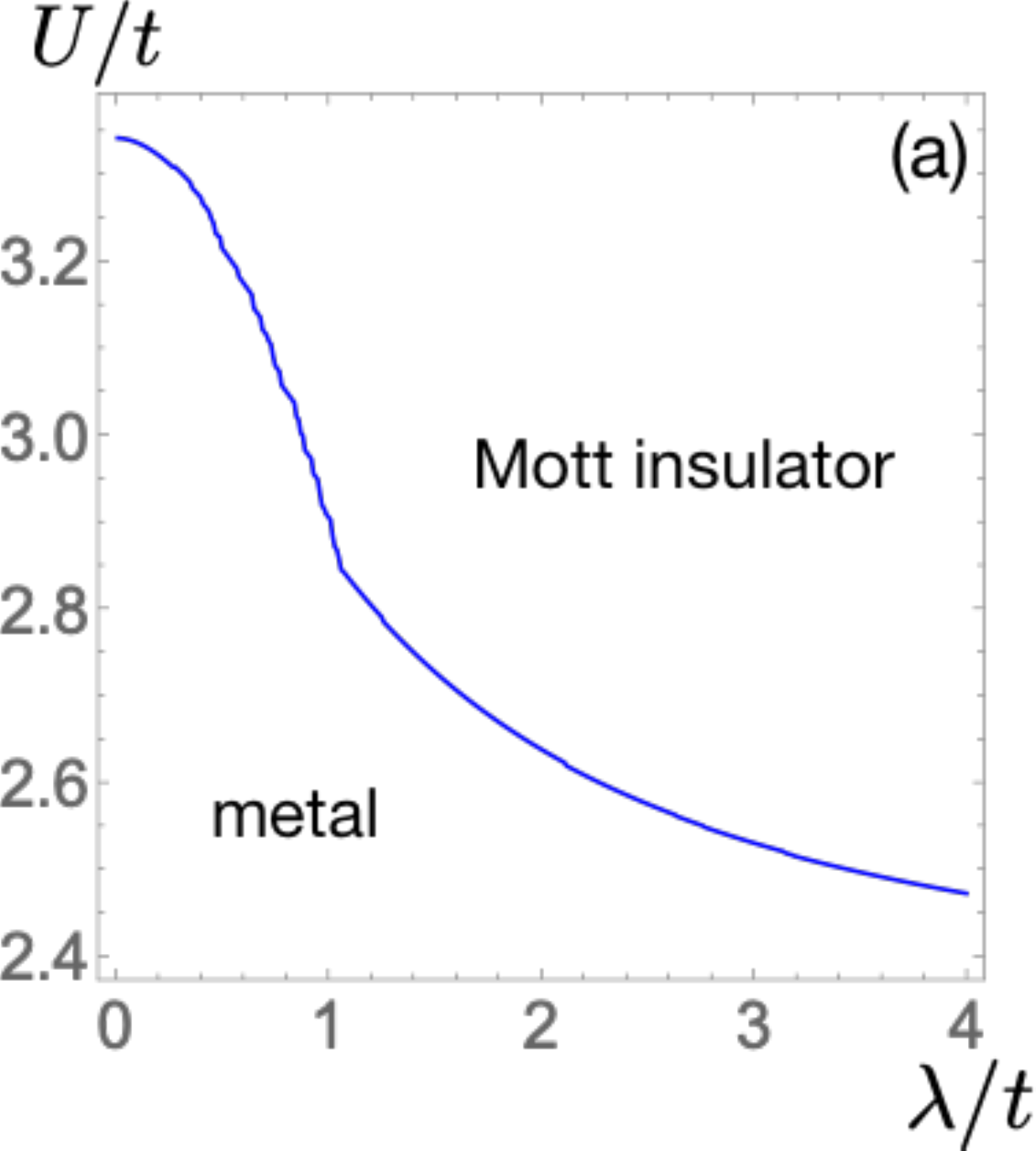}
\includegraphics[width=0.238\textwidth]{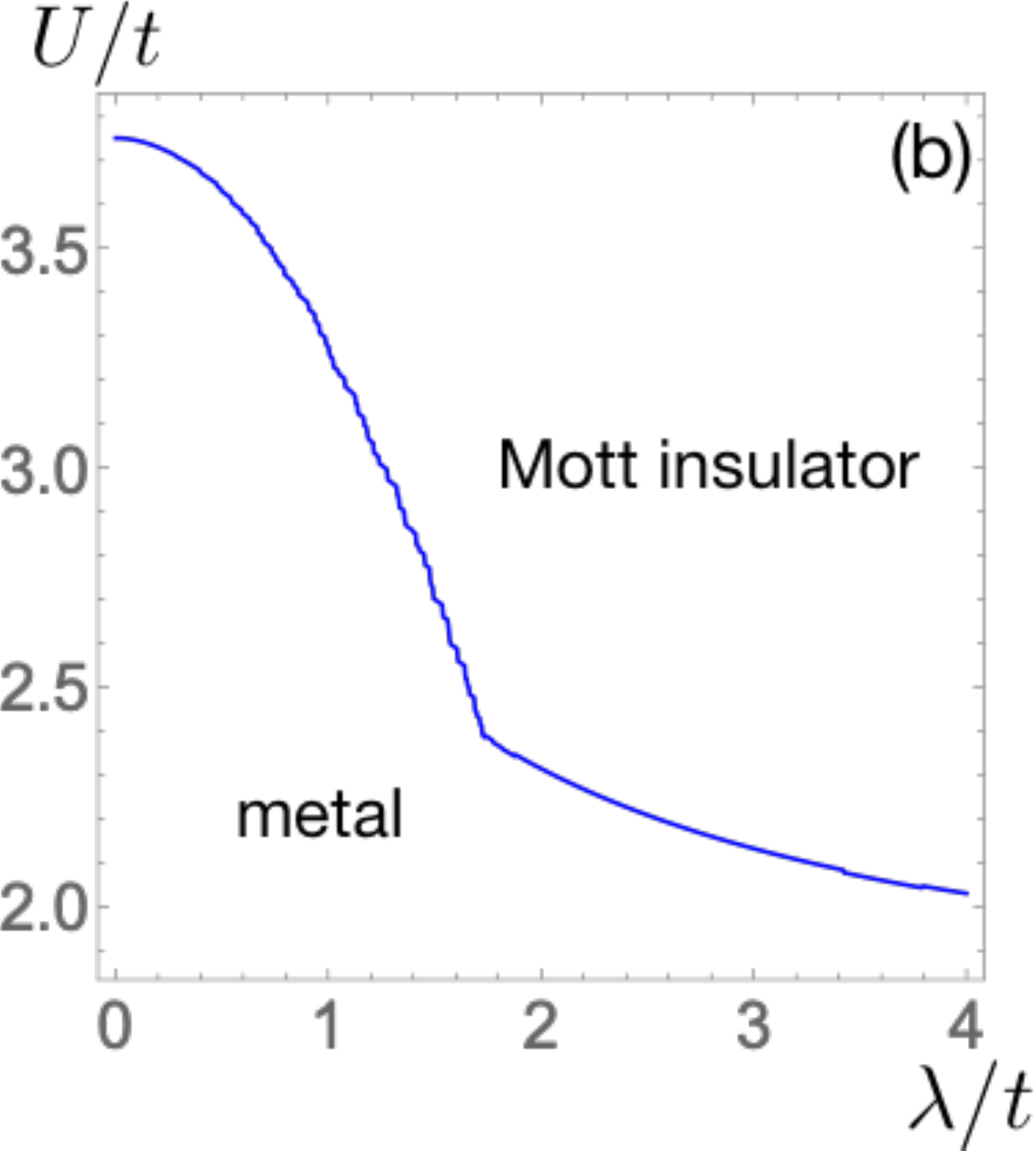}
\includegraphics[width=0.238\textwidth]{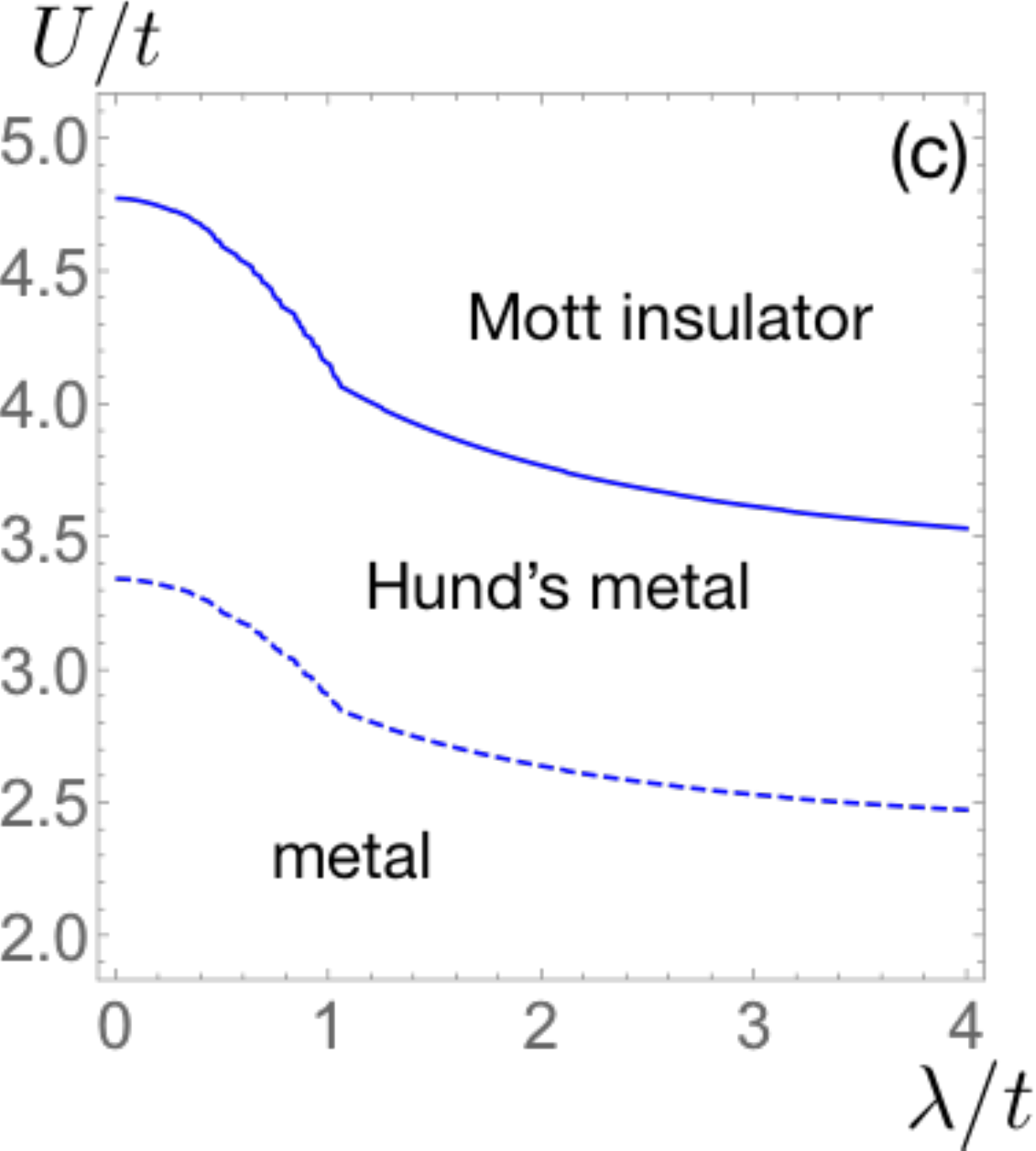}
\includegraphics[width=0.238\textwidth]{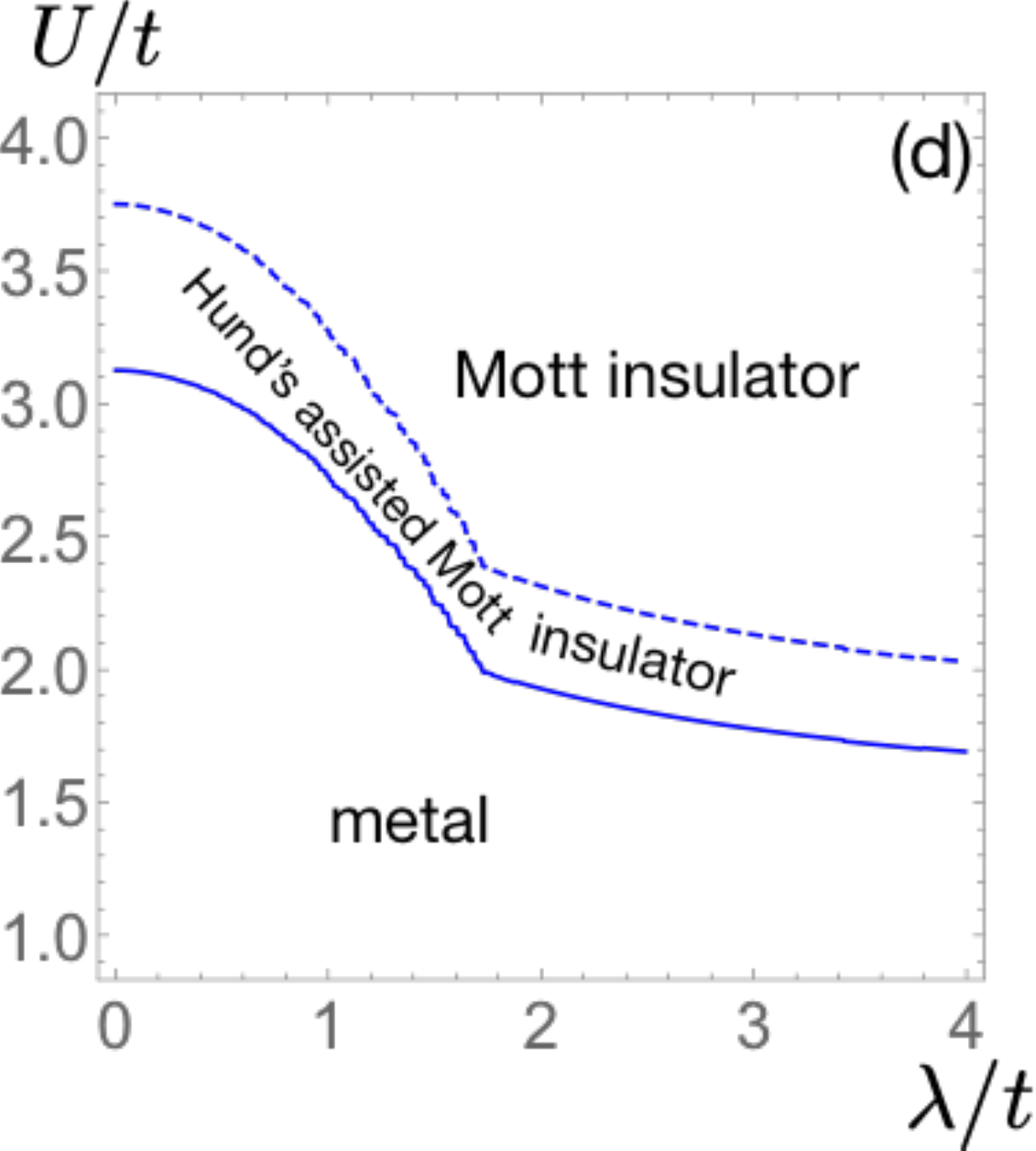}
\caption{(a) and (b) are the Mott transition phase diagrams for the $d^2$ 
and $d^3$ electron configuration in the absence of $J$-interactions, respectively.
(c) and (d) are the Mott transition phase diagrams after taking into account of the renormalization of the correlation by the $J$-interactions. In (c) and (d), the dashed 
curves are the phase boundaries from (a) and (b).  
In (c) and (d), we set $J=0.1U$. 
}
\label{fig2}
\end{figure*}

For our purpose, we first take away the $J$-interactions, and decouple the 
extended Hubbard model into the spinon sector ${\mathcal H}_{f}$ and the 
the charge sector ${\mathcal H}_{\theta}$ with 
\begin{eqnarray}
{\mathcal H}_{f} &=& \sum_{\langle ij \rangle} \sum_{m,n}  
\overline{t}_{ij}^{mn} f^\dagger_{im\alpha} f^{\phantom\dagger}_{jn\alpha}   
+ \sum_i \sum_{m,n} \sum_{\mu} \frac{\lambda}{2}  {L}^{\mu}_{mn}  {\sigma}^{\mu}_{\alpha\beta} 
f^\dagger_{im\alpha} f^{\phantom\dagger}_{in\beta}   \nonumber 
\\
&&\quad\quad\quad\quad\quad\quad\quad\,\,\, -\sum_i \sum_m h_i^{}\, f^{\dagger}_{im\alpha} f^{\phantom\dagger}_{im\alpha}  ,
\\
{\mathcal H}_{\theta} &=&  \sum_{\langle ij \rangle} \big( \chi_{ij}^{} e^{i \theta_i - i\theta_j }  + h.c. \big)
+ \sum_i \big( 
\frac{U}{2}{\mathbb L}_i^2 + h_i {\mathbb L}_i 
\big),
\end{eqnarray}
where ${\overline{t}_{ij}^{mn} = \langle e^{i\theta_i - i \theta_j} \rangle t_{ij}^{mn}}$,
${\chi_{ij}^{} = \sum_{m,n}^{} {t}_{ij}^{mn}  \langle f^\dagger_{im\alpha} 
f^{\phantom\dagger}_{jn\alpha}  \rangle }$, $h_i$ is a Lagrangian 
multiplier for each site to enforce the Hilbert space constraint 
such that ${[\sum_m \sum_{\alpha} 
f^\dagger_{im\alpha}f^{\phantom\dagger}_{im\alpha} ] - \bar{n} 
= {\mathbb L}_i}$, and ${\mathbb L}_i$ is an angular momentum variable conjugate to 
the U(1) phase $\theta_i$. Here $\bar{n}$ is the electron occupation number per site, 
and ${\bar{n}=2}$ (3) for the $d^2$ ($d^3$) electron configuration. 
As the translation symmetry is preserved throughout, one expects
that $h_i$ has no site-dependence and ${h_i \equiv h}$. 
Because ${\langle \sum_i {\mathbb L}_i^{} \rangle =0}$, so we expect 
${h=0}$ in the self-consistent mean-field calculation. Moreover, 
due to the translation symmetry and the lattice rotation, ${\chi_{ij} \equiv \chi}$. 
The charge sector model, ${\mathcal H}_{\theta}$, 
behaves more like a boson Hubbard model at integer fillings, and we solve it 
with a coherent state path integral formulation. After integrating out the 
angular momentum ${\mathbb L}$, we obtain 
\begin{eqnarray}
{\mathcal Z}_{\theta} & \simeq &
\int {\mathcal D} \Phi^\dagger {\mathcal D} \Phi  \nonumber \\
&&
\exp \Big[ { - \big[\int_0^{\beta} d\tau \sum_i  \frac{|\partial_\tau \Phi_i|^2 }{2U}
+ \sum_{\langle ij \rangle} \chi ( \Phi^\dagger_i \Phi_j^{} + h.c.)\big] }  \Big], 
\label{path}
\end{eqnarray}
where we have replaced $e^{-i\theta_i}$ with $\Phi_i$, and $|\Phi_i| =1$. 
The uni-modular condition on $\Phi_i$ can be imposed by introducing 
a Lagrangian multiplier, $\Lambda_i$.  
The excitation spectrum of charge boson $\Phi$ can be solved     
via a saddle point approximation
and the uniformity requirement with ${\Lambda_i=\Lambda}$.
We find that the $\Phi$ spectrum is given by
${\Omega_{\boldsymbol k} = \sqrt{2U [\Lambda - 2|\chi|( \cos k_x + \cos k_y)]}}$
where the lattice constant is set to unity. 
When the gap of the spectrum vanishes, the charge boson $\Phi$ is condensed
and the system goes from the Mott insulator to the metallic state. 
We find that the Mott localization occurs at ${\big[U/|\chi|\big]_c = 4.84}$. 
To obtain the actual critical $U/t$ for the Mott transition, 
one further requires the knowledge from the spinon sector to
produce the parameter $\chi$, and this $\chi$ parameter
depends on $ \langle e^{i\theta_i - i \theta_j} \rangle $ 
at the Mott transition. The latter quantity can then be  
directly computed from Eq.~\eqref{path}, and we find that 
$ \langle e^{i\theta_i - i \theta_j} \rangle = \sum_{\boldsymbol k} 
\Omega_{\boldsymbol k}/(8\chi N)$ for the nearest-neighbor bonds 
at the Mott transition where $N$ is the number of lattice sites.  
The parameter $\chi$ can be evaluated from solving the 
spinon Hamiltonian ${\mathcal H}_f$ with
\begin{eqnarray}
\chi &=& \sum_{m,n} t_{ij}^{mn} \langle f^\dagger_{im\alpha} 
f^{\phantom\dagger}_{jn\alpha} \rangle 
=-\frac{t}{N}\sum_{\boldsymbol k} \sum_{m=2,3}
\langle
f^\dagger_{{\boldsymbol k}m\alpha} 
f^{\phantom\dagger}_{{\boldsymbol k}m\alpha} 
\rangle \cos k_x 
\nonumber \\
&=&-\frac{t}{N} \sum_{\boldsymbol k} \sum_{m=2,3;n}
\big| M({\boldsymbol k})_{m\alpha,n\beta}^{}  \big|^2
             \Theta [\epsilon_{\text F}^{} - \epsilon^{}_{n\beta} ({\boldsymbol k})  ]
           \cos k_x .
           \end{eqnarray}
Here using the translation symmetry, we only need to consider 
the $\hat{x}$ direction bond. The spinon Hamiltonian is diagonalized 
by the canonical transformation 
${f_{{\boldsymbol k}m\alpha}=M({\boldsymbol k})_{m\alpha,n\beta} 
d_{{\boldsymbol k}n\beta}}$, where $d_{{\boldsymbol k}n\beta}$
is the spinon eigenmode and the energy is given by 
$\epsilon_{n\beta}({\boldsymbol k})$ with the spinon Fermi energy    
$\epsilon_{\text F}^{}$. For each $\lambda/t$, there is a corresponding $\chi$
parameter from the spinon sector, and thus a corresponding $(U/t)_c$ for 
the Mott transition. Thereby, we are able to construct the 
phase diagram in the $U/t$-$\lambda/t$ plane.

In Fig.~\ref{fig2}(a) and Fig.~\ref{fig2}(b), we depict the phase diagram 
for the $d^2$ and $d^3$ electron configurations, respectively. 
It is shown that, the critical Hubbard $U$-interaction for the 
Mott transition is gradually suppressed as the spin-orbit coupling 
is increased. It turns out out,  both $d^2$ and $d^3$ fillings have qualitatively
similar phase diagrams as expected. The spin-orbit coupling suppresses
the electron bandwidth, and a weaker $U$-interaction would already drive the Mott 
transition~\cite{Witczak_Krempa_2014}. On the other hand, the Hubbard $U$-interaction suppresses the 
bandwidth, which then enhances the effect of the spin-orbit coupling. These
two interpretations provide a physical understanding of the phase diagram. 
In the right region of the Mott insulating phases, the system should be 
more appropriately quoted as a {\sl relativistic Mott insulator} to reflect the strong
spin-orbit coupling. 
Likewise, In the right region of the metallic phase, the system is better to be 
quoted as a {\sl spin-orbit-coupled metal}~\cite{PhysRevLett.115.026401}.

We now include the effect of the $J$-interactions. As we have previously explained, the 
proper treatment of the $J$-interactions requires the knowledge of the ground state
on the Mott side. We do not intended to address the actual ground state on the Mott side,
and thus, we tend to consider the $J$-interactions in a qualitative manner.  In the strong Mott regime
with decoupled atoms, things can be understood in both qualitatively and quantitatively. 
It is shown that~\cite{Georges_2013,medici2017hunds}, 
the renormalized effective correlation $U_{r}$ can be obtained by calculating
the energy cost for changing the valence charge of two neighboring ions from their original 
electron occupation, {\sl i.e.} transferring one electron from one site to the other. When the 
electron occupation is not at half-filling (i.e. not occupying each orbital with one electron),
the system can gain energies from the Hund's coupling and the inter-orbital interaction, 
and the renormalized correlation is ${U_{\text{r}} = U -3J}$. When the electron occupation 
is at half-filling, transferring electrons would automatically introduce the double electron 
occupation on a single orbital and thus increase the correlation energy. The renormalized 
correlation in this case is ${U_{\text{r}} = U + 2J}$ for our $d^3$ configuration. 
In Fig.~\ref{fig2}(c) and Fig.~\ref{fig2}(d), we depict the new phase diagrams 
after taking into account the renormalized correlation. 
In Fig.~\ref{fig2}(c) for the $d^2$ configuration, 
the effective correlation is reduced by the Hund's coupling,
and thus a large region that was insulating in Fig.~\ref{fig2}(a) 
becomes metallic. This region is nothing but 
Hund's metal. On the right part inside this region, as the strong
spin-orbit coupling is involved, it should be quoted 
as a {\sl spin-orbit-coupled Hund's metal}. 
It turns out that, the $5d$ compound BaOsO$_3$ was 
recently proposed as a spin-orbit-coupled Hund's metal~\cite{bramberger2020baoso3}.
In Fig.~\ref{fig2}(d), for the $d^3$ configuration, 
the effectively correlation is enhanced, and a large metallic region
in Fig.~\ref{fig2}(b) is converted into Mott insulators. This 
is a Hund's assisted Mott insulator, and the right region of it
should then be called a ``Hund's assisted relativistic Mott insulator''.

The interplay between the Hund's coupling and the spin-orbit coupling
persists even in the strong Mott regime. For the $d^2$ configuration
on the $t_{2g}$ shell, if one considers the Hund's coupling first, then one 
arrives with a ${S=1}$ local moment with a three-fold orbital degeneracy for 
the orbital configuration that functions as an effective angular momentum
${L=1}$. Once the spin-orbit coupling is considered, a ${J=2}$ local 
moment is obtained with the ${J=1}$ and ${J=0}$ states as the excited levels~\cite{PhysRevB.84.094420}. 
Another perspective is to first consider the spin-orbit coupling on the single electron 
level and then incorporate the Hund's coupling on top of the spin-orbit energy levels . 
Recent theories in Refs.~\onlinecite{PhysRevB.101.155118,PhysRevLett.124.087206}
noticed that the five-fold degeneracy of 
the ${J=2}$ moment is not protected by the cubic point group symmetry and 
further splitting should be considered. 
For the $d^3$ configuration, the Hund's coupling leads to a total ${S=3/2}$ local moment, 
and the orbital sector is a singlet. The spin-orbit coupling is inactive. If the spin-orbit
coupling is considered first, however, the three electrons would occupy the four-fold degenerate
${J=3/2}$ quadruplets, and four-fold degenerate local states with the spin-orbit entanglement are 
obtained~\cite{PhysRevB.84.094420,PhysRevLett.112.167203}.

\emph{Discussion.}---In our illustrative study of the Hund's coupling and spin-orbit coupling
in the correlated materials, we only considered the $d^2$ and $d^3$ configurations, and the 
uncorrelated or weakly correlated regimes in our examples are all metallic. 
In reality, it could happen that the uncorrelated or weakly correlated regime is a band insulator. 
In that case, the topological aspect of the band structure should be considered. As the spin-orbit
coupling is involved, whether the band insulator is a topological insulator or not is an interesting
and relevant question. Just like the metallic behavior can be driven by the Hund's coupling, 
the candidate topological band insulator if exists is an example of  
Hund's topological insulator. Likewise, one could use Hund's coupling to enlarge the 
region of other topological matter at the single-particle level. 

The parent state of the Fe-based superconductors is often a correlated metal.
As this metal was interpreted as a Hund's metal, the resulting superconductor was 
then proposed as a Hund's superconductor~\cite{PhysRevLett.121.187003,2020arXiv201111475M}. 
The spin-orbit coupling was recently invoked for the Fe-based superconductors~\cite{PhysRevX.7.021025}, 
and various topological features such as Dirac band touching and majorana physics
were proposed~\cite{PhysRevLett.117.047001,Hao_2018}. 
It can be a good chance to include both Hund's coupling and the
spin-orbit coupling together in the future work. 

To conclude, the spin-orbit coupling and the Hund's coupling have opposite effects on the 
electron correlation for the electron occupation off from the half filling of all orbitals.
 At the half filling, the spin-orbit coupling and the Hund's coupling are found to enhance 
 the electron correlation. In real materials, often the spin-orbit coupling in the systems with 
 heavy ions should be seriously considered, and the Hund's coupling is unavoidable almost for 
 any material with electron correlations.

\emph{Acknowledgments.}---This work is supported by research funds 
from the Ministry of Science and Technology of China with grant 
No.~2018YFE0103200, No.~2016YFA0301001 and No.~2016YFA0300500, 
by Shanghai Municipal Science and Technology Major Project with 
Grant No.~2019SHZDZX04, and from the Research Grants Council 
of Hong Kong with General Research Fund Grant No.~17303819 and 
No.~17306520.

\bibliography{refDec.bib}

\end{document}